\documentclass[conference]{IEEEtran}
\IEEEoverridecommandlockouts
\usepackage[utf8]{inputenc}
\usepackage{amsmath,amsfonts,amssymb}
\usepackage[ruled,vlined,algo2e]{algorithm2e}
\usepackage{amsthm}

\newtheorem{lemma}{Lemma}
\newtheorem{corollary}{Corollary}

\usepackage{graphicx}
\usepackage[english]{babel}
\usepackage{hyperref}

\title{Identifying Systems with Symmetries using Equivariant Autoregressive Reservoir Computers}

\author{
    \IEEEauthorblockN{
        Fredy Vides\IEEEauthorrefmark{1}\IEEEauthorrefmark{2},
        Idelfonso B. R. Nogueira\IEEEauthorrefmark{3},
        Gabriela Lopez Gutierrez\IEEEauthorrefmark{2},
        Lendy Banegas\IEEEauthorrefmark{2} and
        Evelyn Flores\IEEEauthorrefmark{2}
    }
    
    \IEEEauthorblockA{\IEEEauthorrefmark{1} Department of Applied Mathematics, UNAH, Tegucigalpa, Honduras.\\(fredy.vides@unah.edu.hn)\\
    \IEEEauthorblockA{\IEEEauthorrefmark{2}Department of Statistics and Research, CNBS, Tegucigalpa, Honduras.\\(fredy.vides@cnbs.gob.hn; gabriela.lopez@cnbs.gob.hn; lendy.banegas@cnbs.gob.hn; evelyn.flores@cnbs.gob.hn)\\}
    }
    \IEEEauthorblockA{\IEEEauthorrefmark{3}Department of Chemical Engineering, Norwegian University of Science and Technology. \\(idelfonso.b.d.r.nogueira@ntnu.no)}
}

\begin{document}
\maketitle
\thispagestyle{plain}
\pagestyle{plain}

\begin{abstract}
The investigation reported in this document focuses on identifying systems with symmetries using equivariant autoregressive reservoir computers. General results in structured matrix approximation theory are presented, exploring a two-fold approach. Firstly, a comprehensive examination of generic symmetry-preserving nonlinear time delay embedding is conducted. This involves analyzing time series data sampled from an equivariant system under study. Secondly, sparse least-squares methods are applied to discern approximate representations of the output coupling matrices. These matrices play a critical role in determining the nonlinear autoregressive representation of an equivariant system. The structural characteristics of these matrices are dictated by the set of symmetries inherent in the system. The document outlines prototypical algorithms derived from the described techniques, offering insight into their practical applications. Emphasis is placed on the significant improvement on structured identification precision when compared to classical reservoir computing methods for the simulation of equivariant dynamical systems.
\end{abstract}


\section{Introduction}

In the dynamic realm of computational science, reservoir computing has emerged as a formidable approach for system identification and dynamic modeling. Notably, this machine learning algorithm excels in processing information from dynamical systems using observed time-series data, primarily due to its minimal training data and computational resource requirements. Yet, the traditional reservoir computing model, reliant on randomly sampled matrices for its underlying recurrent neural network, faces challenges from the numerous metaparameters needing optimization. Recent strides, particularly in nonlinear vector autoregression (NVAR), have marked a significant evolution in reservoir computing by offering efficient alternatives requiring shorter training datasets and less training time \cite{Gauthier2021}.

Despite advances, reservoir computing studies often lack integration between theoretical foundations, algorithmic strategies, and practical applications, often focusing on isolated components. Although reservoir computers (RC) provide distinct advantages over deep learning alternatives in time series analysis and system identification, particularly due to their inherently shallow architecture, which allows them to be viewed as a form of shallow neural network, their implementation often remains overly general. This lack of specificity hinders their adaptability to real-world applications, slowing their practical deployment in various domains \cite{Yan2024}.

This paper introduces an extended autoregressive reservoir computing strategy, tailored for modeling dynamic systems with inherent symmetries. Our work delves into the theoretical underpinnings and algorithmic strategies crucial for realizing these novel reservoir computing architectures. These architectures not only reflect linear and nonlinear autoregressive vector models but also incorporate symmetry-preserving equivariant matrix approximation methods, thereby offering a nuanced approach to equivariant system identification and modeling. Additionally, we present examples that demonstrate the practical implementation and effectiveness of our Equivariant Autoregressive Reservoir Computer (EARC) in real-world simulations.

Our investigation presents as a main contribution: a strategy that bridges structured matrix approximation methods with the robustness of autoregressive models. This joint approach, detailed in Section \ref{sec:structured-approximation}, provides a reliable mathematical solution for identifying  dynamic systems with symmetries accurately and efficiently.

A particularly exciting offshoot of our research is the development of a Python-based toolkit for equivariant reservoir computer model identification. This toolkit embodies the practical application of concepts discussed in Sections \ref{sec:structured-approximation} and \ref{sec:algorithm}, and is readily accessible for scholarly use.

The practical implications of our structured system identification technology are vast and varied, with applications stretching from the nuanced demands of cyber-physical systems simulation \cite{Yuan2019} to toxicity prediction \cite{Cremer2023}.

To demonstrate the practical application of the proposed theory, we present a prototype algorithm for the computation of equivariant autoregressive reservoir computers, rooted in the methodologies expounded upon in Section \ref{sec:structured-approximation}. This is further brought to life in Section \ref{sec:algorithm}, where we outline the algorithm's architecture and functionality.

In Section \ref{sec:experiments}, we present two computational experiments. The first experiment is focused on the identification of a Hamiltonian system that is characterized by its inherent symmetries. The second experiment applies the symmetry-preserving system identification methods discussed in our study to the realm of finance. Here, we simulate the {\em "representation ranking"} of five financial institutions within a given market or portfolio. Inspired by dynamic ecological models for competitive species, this approach assigns a score ranging from $0$ to $1$ to each institution.

Finally, section \ref{sec:experiments} presents a comparative analysis of error metrics, focusing on the root mean squared error (RMSE) and the difference in equivariant measure (\(\Delta_{EM}\)) between the EARC and a next-generation Reservoir Computer (NGRC). The results, summarized in Table \ref{tb:errors}, highlight a clear and consistent advantage of the EARC method across all error metrics. These findings underscore a significant improvement on the structured identification precision of the EARC method over the NGRC method.

\section{Preliminaries and Notation}  \label{sec:pre}

The symbols $\mathbb{R}^+$ and $\mathbb{Z}^+$ will be used to denote the positive real numbers and positive integers, respectively. For any pair $p,n\in \mathbb{Z}^+$ the expression $d_p(n)$ will denote the positive integer $d_p(n)=n(n^{p}-1)/(n-1)+1$. For any finite group $G$, the expression $|G|$ will be used to denote the order of $G$, that is, the number of elements of $G$. The symbol $\mathbf{1}_n$ will be used to denote the vector in $\mathbb{R}^n$ with all of its coordinates equal to $1$.\\

Given a vector time series $\Sigma=\{x_{t}\}_{t\geq 1}\subset \mathbb{R}^n$, a positive integer $L$ and any $t\geq L$, we will write $\mathbf{x}_L(t)$ to denote the vector:
\[
\mathbf{x}_L(t)=\begin{bmatrix}
\mathbf{x}^{(1)}_{L}(t)^\top & \mathbf{x}^{(2)}_{L}(t)^\top & \cdots &  \mathbf{x}^{(n)}_L(t)^\top
\end{bmatrix}^\top \in \mathbb{R}^{nL},
\]
with 
\[
\mathbf{x}^{(j)}_{L}(t)=\begin{bmatrix}
x^{(j)}_{t-L+1} & x^{(j)}_{t-L+2} & \cdots & x^{(j)}_{t-1}  &  x^{(j)}_{t}
\end{bmatrix}^\top \in \mathbb{R}^{L}.
\]
for $1\leq j\leq n$, where $x^{(j)}_{s}$ denotes the scalar $j$-component of each element $x_s$ in the vector time series $\Sigma$, for $s\geq 1$.\\

The identity matrix in $\mathbb{R}^{n\times n}$ will be denoted by $I_n$, and we will write $\hat{e}_{jn}$ to denote the matrices in $\mathbb{R}^{n\times 1}$ representing the canonical basis of $\mathbb{R}^{n}$ (each $\hat{e}_{jn}$ corresponds to the $j$-column of $I_n$). Given a matrix $X \in \mathbb{R}^{m\times n}$, the expression $\|X\|_F$ will denote the Frobenius norm of $X$ determined by the following expression.\\
\[
\|X\|_F:=\sqrt{\sum_{i=1}^m\sum_{j=1}^n x_{ij}^2}
\]

For any integer $n>0$, in this article, we will identify the vectors in $\mathbb{R}^n$ with column matrices in $\mathbb{R}^{n\times 1}$.\\

Given a matrix $A\in \mathbb{R}^{m\times n}$, we write $\mathrm{vec}(A)$ to denote the column vector obtained by stacking the columns of $A$ according to the following expression
\[
\mathrm{vec}(A)=\begin{bmatrix}
    \mathbf{a}[1]^\top & \cdots & \mathbf{a}[n]^\top
\end{bmatrix}^\top
\]
where $\mathbf{a}[j]$ denotes the $j$-column of $A$ for $1\leq j \leq n$. For any $\mathbf{x}\in \mathbb{R}^{mn}$ we will write $\mathrm{vec}^{\dagger}_m(\mathbf{x})$ to denote the operation such that 
$\mathrm{vec}_m^{\dagger}(\mathrm{vec}(A)) = A$ for any $A\in \mathbb{R}^{m\times n}$.

Let $A\in \mathbb{R}^{m\times n}$, $B\in \mathbb{R}^{p\times q}$, the Kronecker tensor product $A\otimes B \in \mathbb{R}^{mp\times nq}$ is determined by the following operation.
\[
A\otimes B = \begin{bmatrix}
a_{11}B & \cdots & a_{1n}B\\
\vdots & \ddots & \vdots\\
a_{m1}B & \cdots & a_{mn}B
\end{bmatrix}
\]
For any integer $p>0$ and any matrix $X\in \mathbb{R}^{m\times n}$, we will write $X^{\otimes p}$ to denote the operation determined by the following expression.
\[
X^{\otimes p} =\left\{
\begin{array}{ll}
X&, p=1\\
X\otimes X^{\otimes (p-1)}&, p\geq 2
\end{array}
\right.
\]
We will also use the symbol $\Pi_p$ to denote the mapping $\Pi_p:\mathbb{R}^n\to \mathbb{R}^{n^p}$ that is determined by the expression  $\Pi_p(x):=x^{\otimes p}$, for each $x\in \mathbb{R}^n$.
Given a list $A_1,A_2,\ldots,A_m$ such that for $1\leq j\leq m$, $A_j\in \mathbb{R}^{n_j\times n_j}$ for some integer $n_j>0$. The expression $A_1\oplus A_2 \oplus \cdots \oplus A_m$ will denote the block diagonal matrix
\[
A_1\oplus A_2 \oplus \cdots \oplus A_m=\begin{bmatrix}
A_1 & & & \\
& A_2 & &\\
& & \ddots & \\
& & & A_m
\end{bmatrix},
\]
where the zero matrix blocks have been omitted. Given two vectors $\mathbf{x}=[x_j]$, $\mathbf{y}=[y_j]$ in $\mathbb{R}^m$, we will write $\mathbf{x}\odot \mathbf{y}$ to denote the operation corresponding to their Hadamard product $\mathbf{x}\odot \mathbf{y} =\begin{bmatrix}
    x_{j}y_{j}
\end{bmatrix}\in \mathbb{R}^m$. The group of orthogonal matrices in $\mathbb{R}^{n\times n}$ will be denoted by $\mathbb{O}(n)$ in this study.

\section{Approximate Transition Map Identification} \label{sec:structured-approximation}
Let us consider discrete-time dynamical systems determined by the pair $(\hat{\boldsymbol{\Sigma}},\mathcal{T})$ with $\hat{\boldsymbol{\Sigma}}\subset \mathbb{R}^n$, where $\mathcal{T}:\hat{\boldsymbol{\Sigma}}\to \hat{\boldsymbol{\Sigma}}$ is a transition map. Given a finite group $G$ and some matrix representation \cite[Definition 3.1.1]{Steinberg2011RepresentationTO} $\boldsymbol{\pi}:G\to G_{\boldsymbol{\pi}}\subset \mathbb{O}(n),g\mapsto g_{\boldsymbol{\pi}}$, the system $(\hat{\boldsymbol{\Sigma}},\mathcal{T})$ is said to be $G$-equivariant with respect to $\boldsymbol{\pi}$ if:
\begin{equation}
g_{\boldsymbol{\pi}}\mathcal{T}(x)=\mathcal{T}(g_{\boldsymbol{\pi}} x)
\label{eq:equiv_constraints_for_EARC}
\end{equation}
for each $x\in \hat{\boldsymbol{\Sigma}}$ and each $g_{\boldsymbol{\pi}}\in \boldsymbol{\pi}(G)$. When it is clear from the context, an explicit reference to $\boldsymbol{\pi}$ may be omitted.\\

Given $\epsilon>0$, some matrix representation $G_{\boldsymbol{\pi}}\subset \mathbb{O}(n)$ of a finite group $G$, and an orbit of a $G$-equivariant discrete-time system to be identified that can be represented by a vector times series $\Sigma=\{x_{t}\}_{t\geq 1}\subset \mathbb{R}^{n}$. We will study the problem of identifying a (generally nonlinear) map $\hat{\mathcal{T}}:\mathbb{R}^{n}\to \mathbb{R}^{n}$ such that
\begin{align}
gx_{t+1}=\hat{\mathcal{T}}(gx_t),
\label{eq:approx_rep_inequality}
\end{align}
for each $1\leq t\leq \tau$, each $g\in G_{\boldsymbol{\pi}}$ and some prescribed $\tau>0$, with $x_{t+1}=\hat{\mathcal{T}}(x_t)$ for each $t\geq 1$.

\subsection{Equivariant reservoir computers for approximate transition map identification}
\label{section:EARC_theory_and_methods}

When for a matrix representation $\boldsymbol{\pi}:G\to \mathbb{O}(n)$ of some given finite group $G$, we consider the time series data $\Sigma\subset \mathbb{R}^n$ corresponding to an orbit determined by the difference equation 
\begin{equation}
x_{t+1} = \mathcal{T}(x_t),
\label{eq:original_system_dynamics_diff_eq}
\end{equation}
for some $G$-equivariant discrete-time system $(\hat{\boldsymbol{\Sigma}},\mathcal{T})$ with respect to $\boldsymbol{\pi}$ to be identified. One may need to preprocess the time series data before proceeding with the approximate representation of a suitable transition map. For this purpose, given some prescribed integer $L>0$, one can consider the time series $\mathcal{D}_L(\Sigma)$ determined by the expression:
\[
\mathcal{D}_L(\Sigma)=\{\mathbf{x}_L(t)\}_{t\geq L}
\]
For the dilated time series $\mathcal{D}_L(\Sigma)$, the previously considered recurrence relation $x_{t+1}=\mathcal{T}(x_t)$, $t\geq 1$, induces the following difference equations
\begin{align}
\mathbf{x}_L(t+1) = \tilde{\mathcal{T}}(\mathbf{x}_L(t)),
\label{eq:dilated_dynamics_diff_eq}
\end{align}
for $t\geq L$. Where $\tilde{\mathcal{T}}$ is some transition map to be identified such that
\begin{align}
(\boldsymbol{\pi}(g)\otimes I_L)\mathbf{x}_L(t+1) = \tilde{\mathcal{T}}\left((\boldsymbol{\pi}(g)\otimes I_L)\mathbf{x}_L(t)\right),
\label{eq:dilated_equiv_constraints}
\end{align}
for each $g\in G$.

For any $p\geq 1$, let us consider the map $\eth_p:\mathbb{R}^n\to \mathbb{R}^{d_p(n)}$ for $d_p(n)=n(n^{p}-1)/(n-1)+1$, which is determined by the expression: 
\begin{equation}
    \eth_p(x):=
\begin{bmatrix}
\Pi_1(x)\\
\Pi_2(x)\\
\vdots\\
\Pi_p(x)\\
1
\end{bmatrix}=
\begin{bmatrix}
x^{\otimes 1}\\
x^{\otimes 2}\\
\vdots\\
x^{\otimes p}\\
1
\end{bmatrix}
\label{eq:eth_map}
\end{equation}
Here, the number $p$ will be called the order of the embedding map $\eth_p$. Given integers $p,L>0$, an orbit $\Sigma=\{x_t\}_{t\geq 1}\subset \mathbb{R}^n$ of an equivariant system with a finite symmetry group represented by a set of orthogonal matrices $G\subset \mathbb{R}^{n\times n}$. For a finite sample $\Sigma_T = \{x_t\}_ {t=1}^T\subset \Sigma$, let us  consider the matrices:
\begin{align}
\mathbf{H}^{(0,p)}_{L}(\Sigma_T)&=\begin{bmatrix}
\eth_p(\mathbf{x}_L(L)) & \cdots & \eth_p(\mathbf{x}_L(T-1))
\end{bmatrix}\label{eq:structured_data_matrices}\\
\mathbf{H}^{(1)}_{L}(\Sigma_T)&=\begin{bmatrix}
\mathbf{x}_L(L+1) & \cdots & \mathbf{x}_L(T)
\end{bmatrix}\nonumber
\end{align}

The mapping identification mechanism used in this study for dilated systems of the form \eqref{eq:dilated_dynamics_diff_eq}, will be described by the expression:
\begin{equation}
\hat{\mathcal{T}}(\mathbf{x}_L(t))=W\eth_p(\mathbf{x}_L(t)), \:\: t\geq L,
\label{eq:evolution_op_id}
\end{equation}
for some matrix $W=\hat{W}R_{p,L}(n)\in \mathbb{R}^{n\times d_p(n)}$ to be partially determined. Building on matrix theoretic techniques and ideas presented in \cite{doi:10.1137/090779966}, \cite{10.1063/5.0071154} and \cite{vides2023dynamic}, the matrix $\hat{W}$ in \eqref{eq:evolution_op_id} can be estimated by approximately solving the matrix equation
\begin{equation}
\hat{W}(R_{p,L}(n)\mathbf{H}^{(0,p)}_{L}(\Sigma_T)) = \mathbf{H}^{(1)}_{L}(\Sigma_T).
\label{eq:evol_matrix_eq}
\end{equation}
Where $R_{p,L}(n)$ is the matrix described by \cite[Theorem III.6]{vides2023dynamic}, and by \eqref{eq:dilated_dynamics_diff_eq} and \eqref{eq:dilated_equiv_constraints} $\hat{W}$ shall belong to the linear space of matrices that solve equations of the form:
\begin{align}
\|g_j\otimes I_L X-XR_{p,L}(n)G_j\|_F=0, 1\leq j\leq |\mathbf{G}_{\boldsymbol {\pi}}|
\label{eq:structured_main_matrix_eq}
\end{align}
and where for each $1\leq j\leq |\mathbf{G}_{\boldsymbol{\pi}}|$ the matrix $G_j$ satisfies the condition 
\begin{align}
\eth_p(g_j\otimes I_L\mathbf{x})=G_j\eth_p(\mathbf{x})
\label{eq:structured_secondary_matrix_eq}
\end{align}
for any $\mathbf{x}\in \mathbb{R}^{nL}$.
The devices described by \eqref{eq:evolution_op_id} are called equivariant autoregressive reservoir computers (EARC) in this paper. 

\subsection{Structured output coupling matrix identification}
The solvability of the matrix identification problems described by equations \eqref{eq:structured_main_matrix_eq} and \eqref{eq:structured_secondary_matrix_eq} will be studied from a structured matrix analysis perspective.

\begin{lemma}
\label{lema:right-action-representation}
Given an integer $p\geq 1$, and a finite group representation $\mathbf{G}_{\boldsymbol{\pi}}=\{g_1,\ldots,g_N\}\subset \mathbb{O}(n)$. For each $1\leq j\leq N=|\mathbf{G}_{\boldsymbol{\pi}}|$, the matrix $G_j$ satisfies the condition 
\begin{align*}
\eth_p(g_j\otimes I_L\mathbf{x})=G_j\eth_p(\mathbf{x})
\end{align*}
for any $\mathbf{x}\in \mathbb{R}^{nL}$, is determined by the expression
\begin{equation}
G_j:=(g_j\otimes I_L)\oplus (g_j\otimes I_L)^{\otimes 2}\oplus \cdots \oplus (g_j\otimes I_L)^{\otimes p}\oplus 1.
    \label{eq:right-action-def}
\end{equation}
\end{lemma}
\begin{proof}
By iterating on \cite[Lemma 2.1]{zhan2013matrix} we will have that for any $\mathbf{x}\in \mathbb{R}^{nL}$ and any integers $k\geq 1$ and $1\leq j\leq N$:
\begin{align*}
    ((g_j\otimes I_L)\mathbf{x})^{\otimes k}&=(g_j\otimes I_L)^{\otimes k}\mathbf{x}^{\otimes k}\\
    &=(g_j\otimes I_L)^{\otimes k}\Pi_k(\mathbf{x})
\end{align*}
Consequently, by \eqref{eq:eth_map} it can be seen that if $G_j$ is determined by \eqref{eq:right-action-def} we will have that
\begin{align*}
    \eth_p(g_j\otimes I_L\mathbf{x})=G_j\eth_p(\mathbf{x}).
\end{align*}
This completes the proof.
\end{proof}

Given a finite group representation $\mathbf{G}_{\boldsymbol{\pi}}=\{g_1,\ldots,g_N\}\subset \mathbb{O}(n)$ and a subset of matrix representations $\{\hat{g}_1,\ldots,\hat{g}_r\}\subset \mathbf{G}_{\boldsymbol{\pi}}$ of the generators of $G$, then a basis for the space of solutions to equations \eqref{eq:structured_main_matrix_eq} can be computed by applying the following lemma.

\begin{lemma}
\label{lem:first-equivariance-lemma} 
Given a subset of matrix representations $\{\hat{g}_1$ $,$ $\ldots$ $,$ $\hat{g}_r\}$ $\subset$ $\mathbf{G}_{\boldsymbol{\pi}}$ of the generators of a group of symmetries $G$ of a $G$-equivariant system under study whose output coupling matrix is determined by \eqref{eq:evol_matrix_eq}, then a basis $\{X_1,\ldots,X_M\}$ for the space of solutions to equations \eqref{eq:structured_main_matrix_eq} will be determined by a basis $\{\mathbf{x}_1,\ldots,\mathbf{x}_M\}$ of
\begin{equation}
\ker \left(\sum_{j=1}^r K_j^\top K_j\right)
    \label{eq:symmetries_kernel}
\end{equation}
according to the rule 
\begin{equation}
X_j=\mathrm{vec}_{nL}^{\dagger}(\mathbf{x}_j), 1\leq j\leq M.    
\end{equation}
Where 
\begin{equation}
    K_j:=I_{q_1}\otimes (\hat{g}_j\otimes I_L)-(G_j^\top R_{p,L}(n)^\top)\otimes I_{q_2}
    \label{eq:Kj-def}
\end{equation}
for each $1\leq j\leq r$ and some suitable integers $q_1,q_2\geq 1$.
\end{lemma}
\begin{proof}
Let $q_1=nL$ and $q_2$ be equal to the number of rows of $R_{p,L}(n)$ determined by \cite[Theorem III.6]{vides2023dynamic}. Since $\hat{g_1},\ldots,\hat{g}_r$ represent generators of the group of symmetries $G$, by \cite[Equation (2.10) in \S2.2]{zhan2013matrix}, \eqref{eq:Kj-def}, and by matrix kernel properties we will have that any matrix solvent $X$ of the matrix equations \eqref{eq:structured_main_matrix_eq} satisfies the following condition.
\begin{equation}
\mathrm{vec}(X)\in  \ker\left(
\begin{bmatrix}
    K_1\\
    \vdots\\
    K_r
\end{bmatrix}
\right)=\ker \left(
\sum_{j=1}^r K_j^\top K_j
\right)
    \label{eq:joint_symmetric_kernel}
\end{equation}
Consequently, given a basis $\{\mathbf{x}_1,\ldots,\mathbf{x}_M\}$ of \eqref{eq:symmetries_kernel} we will have that $\mathrm{vec}(X)\in \mathrm{span}(\{\mathbf{x}_1,\ldots,\mathbf{x}_M\})$. Therefore, $X\in \mathrm{span}(\{\mathrm{vec}_{nL}^{\dagger}(\mathbf{x}_1),\ldots,\mathrm{vec}_{nL}^{\dagger}(\mathbf{x}_M)\})$ and $\{\mathrm{vec}_{nL}^{\dagger}(\mathbf{x}_1)$ $,$ $\ldots$ $,$ $\mathrm{vec}_{nL}^{\dagger}(\mathbf{x}_M)\}$ determines a basis for the space of solutions to \eqref{eq:structured_main_matrix_eq}. This completes the proof.
\end{proof}

\begin{corollary}
Let us consider the matrix equation \eqref{eq:evol_matrix_eq}, and let us set:
\begin{align*}
    \mathbf{H}_0&:=R_{p,L}(n)\mathbf{H}^{(0,p)}_{L}(\Sigma_T),\\
    \mathbf{H}_1&:= \mathbf{H}^{(1)}_{L}(\Sigma_T).
\end{align*}
A structured output coupling matrix $\hat{W}$ that satisfies \eqref{eq:evol_matrix_eq} and \eqref{eq:structured_main_matrix_eq}, can be identified using the expression
\[
\hat{W}:=\sum_{j=1}^m c_j X_j.
\]
Where $\{X_1,\ldots,X_M\}$ is the basis for the space of solutions to equations \eqref{eq:structured_main_matrix_eq} described by Lemma \ref{lem:first-equivariance-lemma}, and the coefficients $c_j$ can be obtained by solving the following linear system of equations
\begin{align}
\begin{bmatrix}
| &  & |\\
    \mathrm{vec}(X_1\mathbf{H}_0) & \cdots & \mathrm{vec}(X_M\mathbf{H}_0)\\
    | &  & |
\end{bmatrix} 
\begin{bmatrix}
    c_1\\
    \vdots\\
    c_M
\end{bmatrix}=
\mathrm{vec}(\mathbf{H}_1).
    \label{eq:main_lSP}
\end{align}
\end{corollary}

\begin{proof}
This is can be verified by applying the operation $\mathrm{vec}$ to both sides of \eqref{eq:evol_matrix_eq}, and is a consequence of linear space bases properties and the definition of the operation $\mathrm{vec}$, as it can be seen from its definition that the operation $\mathrm{vec}$ is linear. This completes the proof.
\end{proof}

\section{Algorithm}
\label{sec:algorithm}

In this section, we focus on the applications of the structured matrix approximation methods presented in \S\ref{sec:structured-approximation}, to reservoir computer models identification for equivariant dynamical systems. More specifically, we propose a prototypical algorithm for general purpose equivariant system identification, that is described by Algorithm \ref{alg:main_AutoRegressor_alg_1}.

\begin{algorithm2e}
\caption{{\bf EARCModel}: EARC model identification}
\label{alg:main_AutoRegressor_alg_1}
\SetAlgoLined
 \KwData{$\Sigma_{T}=\{x_{t}\}_{t=1}^{T}\subset \mathbb{R}^n, \mathbf{G}_{\boldsymbol {\pi}}\subset \mathbb{O}(n)$.}
  \KwResult{$\hat{W},R_{p,L}(n)$}
\begin{itemize}
\item [0:] Choose or estimate the lag value $L$ using auto-correlation function based methods.
\item [1:] Set a tensor order value $p$.
\item[2:] Compute compression matrix $R_{p,L}(n)$ applying Algorithm A.2 in \cite{vides2023dynamic}.
\item[3:] Compute matrices:
\begin{align*}
\mathbf{H}_0 &:= \mathbf{H}^{(0,p)}_{L}(\Sigma_T)\\
\mathbf{H}_1 &:= \mathbf{H}^{(1)}_{L}(\Sigma_T)
\end{align*}
\item[4:] For each $g_j\in \mathbf{G}_{\boldsymbol{\pi}}$, compute the matrix $G_j$ such that $\eth_p(g_j\otimes I_L\mathbf{x})=G_j\eth_p(\mathbf{x})$ for any $\mathbf{x}\in \mathbb{R}^{nL}$.
\item[5:] Compute a basis $\{\hat{X}_1,\ldots,\hat{X}_m\}$ for the space of solutions to equations:
\[
g_j\otimes I_L X-XR_{p,L}(n)G_j=\mathbf{0}_n, 1\leq j\leq |\mathbf{G}_{\boldsymbol {\pi}}|
\]
\item[6:] Compute the coefficients that approximately solve: 
\[
\sum_{j=1}^m c_j\hat{X}_j \left(R_{p,L}(n)\mathbf{H}_0\right) = \mathbf{H}_1
\]
applying Algorithm A.1 in \cite{vides2023dynamic}.
\item [7:] Set $\hat{W} := \sum_{j=1}^m c_j \hat{X}_j$.
\end{itemize}
\KwRet{$\hat{W},R_{p,L}(n)$}
\end{algorithm2e}
\section{Computational Examples}
\label{sec:experiments}

In this section we will present some numerical simulations computed using the {\bf SPORT} toolset available in \cite{FVides_SPORT}, which was developed as part of this project, the toolset consists of a collection of programs written in Python that can be used for sparse identification and numerical simulation of dynamical systems. 

The numerical experiments documented in this section were performed with Python 3.11.5 . All the programs written for synthetic data generation and sparse model identification as part of this project are available at \cite{FVides_SPORT}.

\subsection{Identification of a Hamiltonian system with symmetries using EARC}
Let us consider a Hamiltonian system determined by the following initial value problem.
\begin{align}
&\frac{d q}{dt}=p^3-p,\label{eq:HamiltonianModel}\\
&\frac{d p}{dt}=q^3-q, \nonumber\\
&q(0)=1,p(0)=0\nonumber
\end{align}
As observed in \cite{SINHA20201150} the system \eqref{eq:HamiltonianModel} is $K_4$-equivariant. For the configuration used for this experiment, the matrix representation $\mathbf{G}_{\boldsymbol{\pi}}(K_4)$ of the corresponding group of symmetries $$K_4=\left\langle r_1,r_2\left| r_1^2=r_2^2=\left(r_1r_2\right)^2=e\right.\right\rangle$$ 
is determined by the following assignments.
\begin{align*}
r_1&\mapsto r_{1,\boldsymbol{\rho}} =\begin{bmatrix}
0 & 1\\
1 & 0 
\end{bmatrix}\\
r_2&\mapsto r_{2,\boldsymbol{\rho}}= \begin{bmatrix}
-1 & 0\\
0 & -1
\end{bmatrix}
\end{align*}
The synthetic signals corresponding to the data sample $\Sigma_{600}\subset \mathbb{R}^{3}$ that will be used for system identification have been computed with an explicit time integration scheme, using lsoda from the FORTRAN library odepack via the Python function odeint, with the Python program {\tt HamiltonianSystem.py} in \cite{FVides_SPORT}. The model was trained using an embedding map $\eth_p$ of order $p=3$, with $15\%$ of the synthetic reference data.\\

\begin{center}
\begin{figure}[h]
\includegraphics[scale=.42]{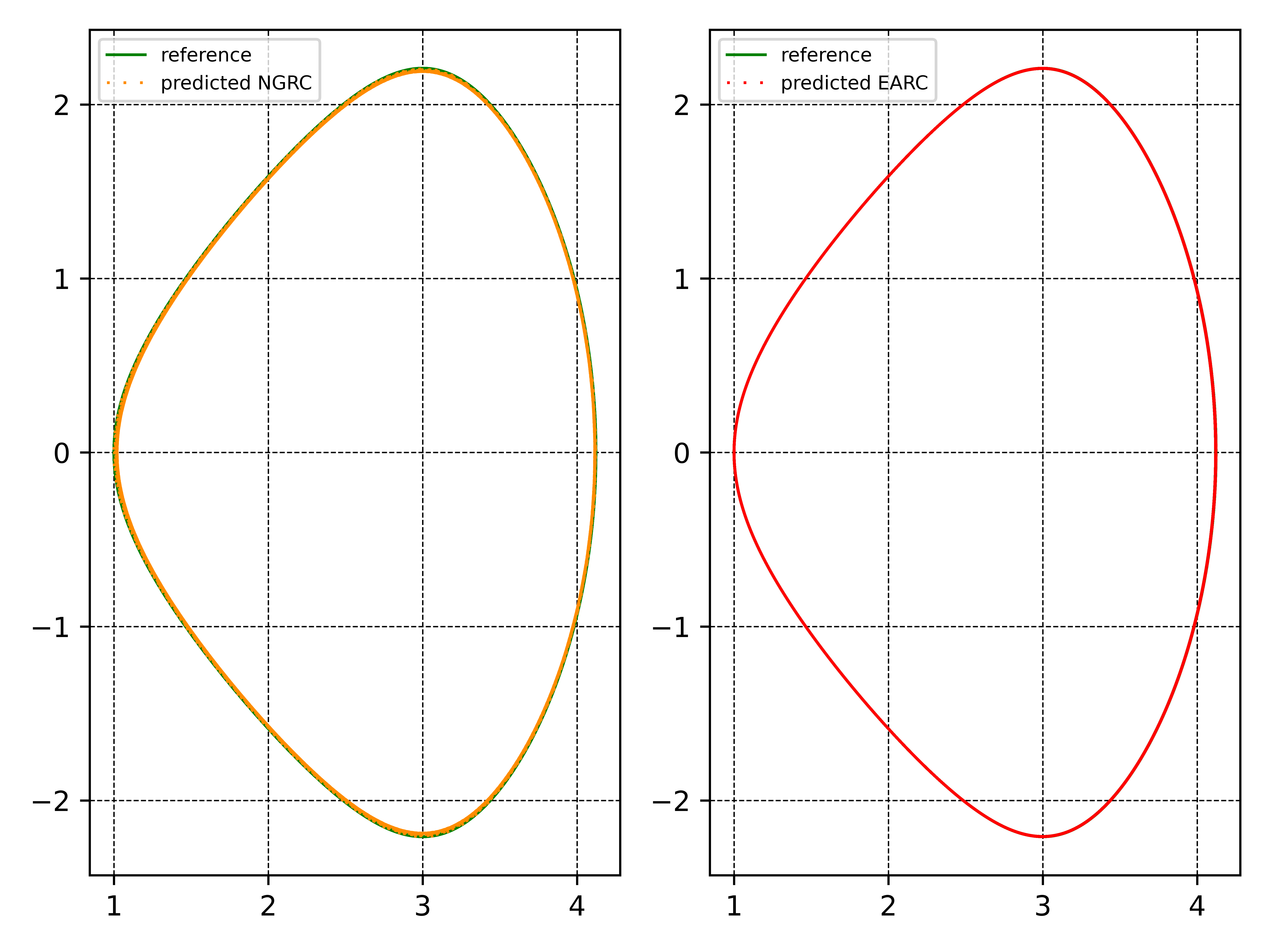}
\caption{$2D$ graphical representation of: reference behavior and predicted behavior NGRC model (left). Reference behavior and predicted behavior EARC model (right).}
\label{fig:Hamiltonian_ID_1}
\end{figure}
\end{center}

The reference synthetic signal data and the corresponding identified signals are illustrated in Figure \ref{fig:Hamiltonian_ID_1}.

For this experiment we have considered the following configuration: a lag value $L=5$, an embedding order $p=3$. To verify that the output coupling matrix $\hat{W}$ identified for this example using Algorithm \ref{alg:main_AutoRegressor_alg_1} satisfies the equivariant matrix constraints \eqref{eq:structured_secondary_matrix_eq}, we can compute the number:
\begin{align*}
    \boldsymbol{\Delta}_{EM}:=\sum_{g_j\in \mathbf{G}_{\boldsymbol{\pi}}(K_4)} \left\|g_j\otimes I_{5}\hat{W}R_{3,5}(2)-\hat{W}R_{3,5}(2)G_j\right\|_F
\end{align*}
Where each $G_j$ is determined by each $g_j$ according to \eqref{eq:right-action-def}. The corresponding structured and none structured errors for this experiment are document in row Exp. 1 of Table \ref{tb:errors}.

The computational setting used for the experiments performed in this section is documented in the Jupyter notebook {\emph{SPORT Simulations}  } under section {\tt COMPUTATIONAL EXAMPLE 1: Identification of a Hamiltonian system with symmetries using EARC} in \cite{FVides_SPORT} that can be used to replicate these experiments.

\subsection{Identification of a financial competition system with symmetries using EARC}
This test case focuses on the dynamic system governing the {\em representation ranking} of financial firms, a quantitative metric ranging from \(0\) to \(1\), which represents each firm's relative participation. A score of \(1\) denotes market dominance, whereas \(0\) indicates an absence of participation. We compute the ranking by normalizing each firm's participation against the maximum observed value in a selected market or portfolio. The characteristics of this model proves useful for regulatory bodies in assessing systemic importance, monitoring competitive dynamics, identifying systemic risks, and implementing measures that enhance financial stability.

\begin{figure}[!ht]
    \centering
    \includegraphics[scale = 0.42]{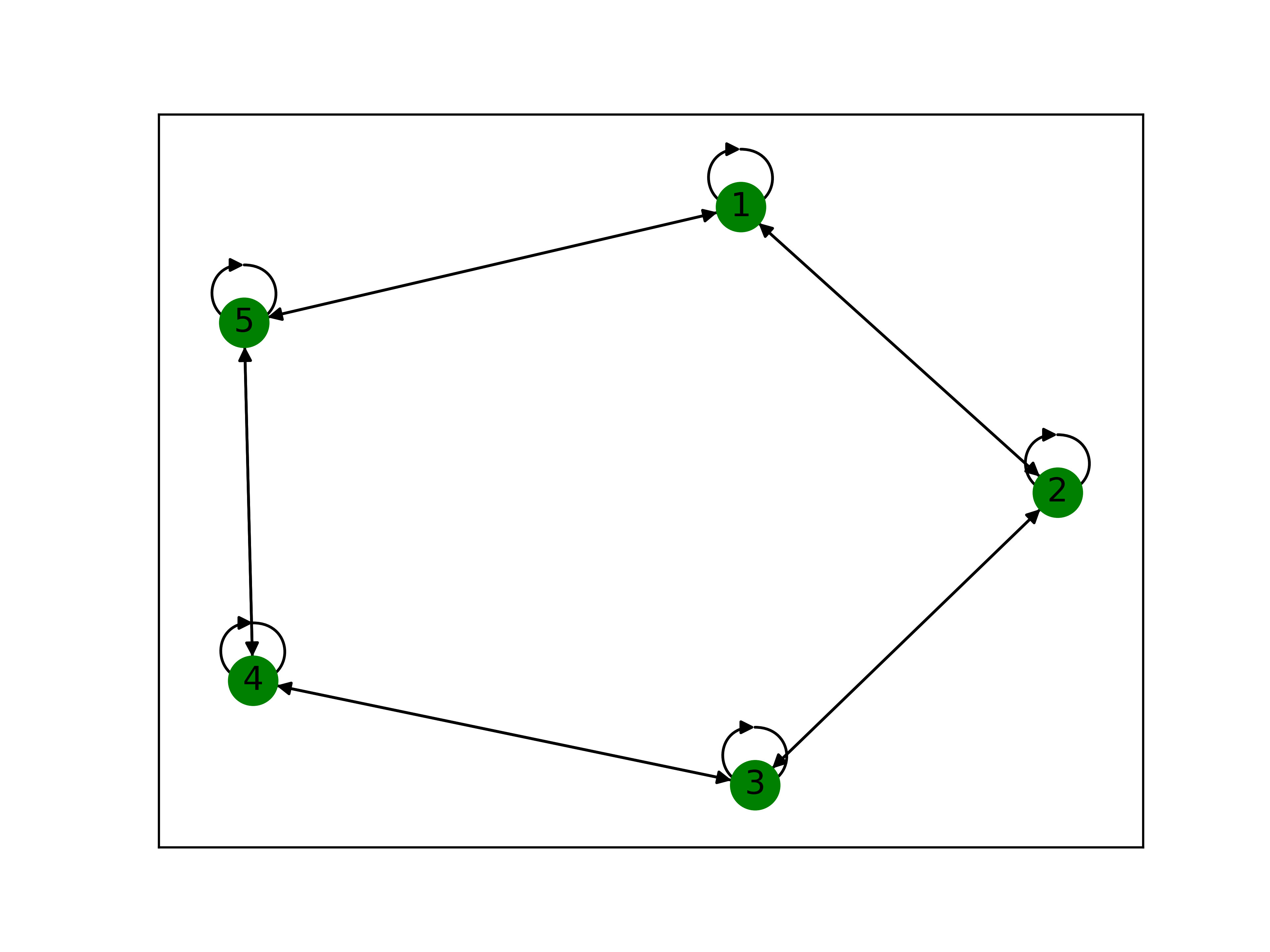}
    \caption{Expected interaction network $\mathcal{N}=(V_\mathcal{N},E_{\mathcal{N}})$.}
    \label{fig:interaction-network}
\end{figure}

Let us consider a discrete-time financial competition system for $5$ banks, characterized by the following recurrence relation:
\begin{align}
\mathbf{p}(t+1)=\mathbf{p}(t)+\mathbf{r}\odot \mathbf{p}(t)\odot(\mathbf{1}_5-\mathbf{N}\mathbf{p}(t))
\label{eq:FiancialModel}
\end{align}
In this model, the $i$-th component $\mathbf{p}(t)[i]$ of each vector $\mathbf{p}(t)$ is an approximate representation for the value at time step $t$, of the discrete-time signal determined by the smoothed evolution of the representation ranking of bank $i$ in a given market or portfolio, as assessed by some regulating body. The matrix $\mathbf{N}$, is defined as:
\begin{align*}
\mathbf{N}& =\begin{bmatrix}
1 &  1.1 & 0 &  0 &  1 \\
 1 & 1 & 1.1 & 0 & 0 \\
 0 & 1 & 1 & 1.1 & 0 \\
 0 & 0 & 1 & 1 & 1.1 \\
 1.1 & 0 & 0 & 1 & 1
\end{bmatrix}
\end{align*}
and represents the interaction between banks of the system under consideration. It corresponds to the expected interaction network $\mathcal{N}=(V_\mathcal{N},E_{\mathcal{N}})$ of the $5$ banks considered in this experiment. The corresponding directed graph is illustrated in Figure \ref{fig:interaction-network}.

When all components of the growth vector $\mathbf{r}$ are equal, the system becomes $\mathbb{Z}_5$-equivariant. For the configuration used for this experiment, $\mathbf{r}=0.376\mathbf{1}_5$ and the matrix representation $\mathbf{G}_{\boldsymbol{\pi}}(\mathbb{Z}_5)$ of the corresponding group of symmetries $\mathbb{Z}_5=\langle r| r^5=e\rangle$ is determined by the following assignment.
\begin{align*}
r&\mapsto r_{\boldsymbol{\rho}} =\begin{bmatrix}
0 & 1 & 0 & 0 & 0\\
0 & 0 & 1 & 0 & 0\\
0 & 0 & 0 & 1 & 0\\
0 & 0 & 0 & 0 & 1\\
1 & 0 & 0 & 0 & 0\\
\end{bmatrix}
\end{align*}

The synthetic signals corresponding to the data sample $\Sigma_{425}\subset \mathbb{R}^{5}$ that will be used for system identification have been computed according to \eqref{eq:FiancialModel}. The model was trained using an embedding map $\eth_p$ of order $p=2$, with less than $7.3\%$ of the synthetic reference data.

\begin{center}
\begin{figure}[!ht]
\includegraphics[scale=.5]{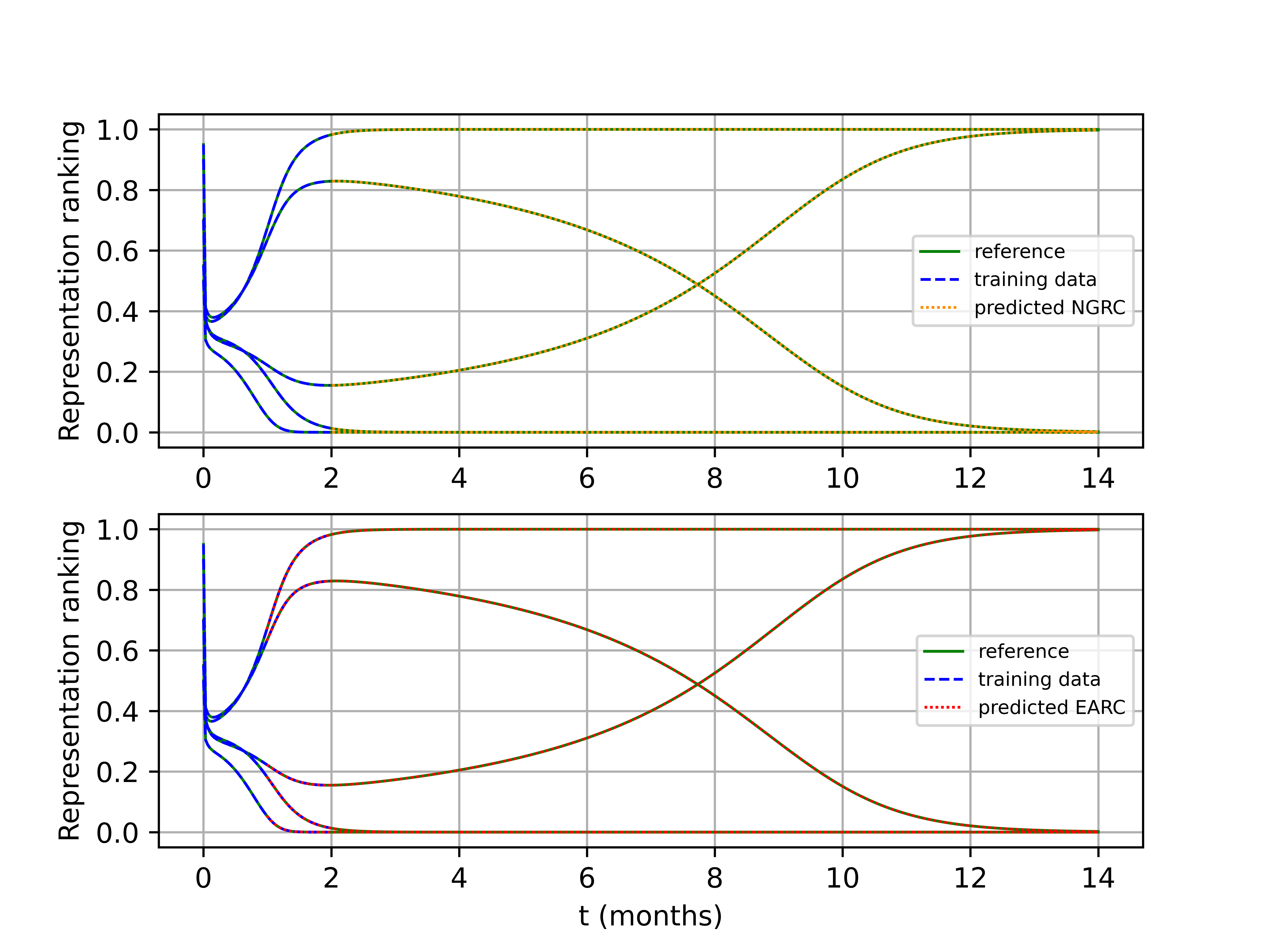}
\caption{Representation ranking evolution: reference behavior data, training data and predicted behavior NGRC model (top). Reference behavior data, training data and predicted behavior NGRC model (bottom).}
\label{fig:Comtetition_ID_1}
\end{figure}
\end{center}

For this experiment, we have considered the following configuration: a lag value $L=1$, and an embedding order $p=2$.The reference synthetic signal data and the corresponding identified signals are illustrated in Figure \ref{fig:Comtetition_ID_1}.

To verify that the output coupling matrix $\hat{W}$ identified for this example using Algorithm \ref{alg:main_AutoRegressor_alg_1} satisfies the equivariant matrix constraints \eqref{eq:structured_secondary_matrix_eq}, we can compute the number:
\begin{align*}
    \boldsymbol{\Delta}_{EM}:=\sum_{g_j\in \mathbf{G}_{\boldsymbol{\pi}}(\mathbb{Z}_5)} \left\|g_j\hat{W}R_{2,1}(5)-\hat{W}R_{2,1}(5)G_j\right\|_F
\end{align*}
Where each $G_j$ is determined by each $g_j$ according to \eqref{eq:right-action-def}. For this experiment, the corresponding structured and none structured errors are documented in row Exp. 2 of Table \ref{tb:errors}.


The computational setting used for the experiments performed in this section is documented in the Jupyter notebook {\emph{SPORT Simulations}  } under section {\tt COMPUTATIONAL EXAMPLE 2: Identification of a financial competition system with symmetries using EARC} in \cite{FVides_SPORT} that can be used to replicate this experiment.

\begin{table}[!hb]
\begin{center}
\caption{Error Metrics}\label{tb:errors}
\begin{tabular}{ccccc}
  & RMSE-S  & RMSE-N &$\Delta_{EM}$-S& $\Delta_{EM}$-N \\\hline
Exp. 1 & $6.58 \times 10^{-5}$ & $1.19$ & $3.76 \times 10^{-15}$ & $3.15$\\
Exp. 2 & $4.07 \times 10^{-13}$ & $3.00 \times 10^{-6}$ & $1.03 \times 10^{-14}$  & $5.63 \times 10^{-7}$\\ \hline
\end{tabular}
\end{center}
\end{table}

\section{Data Availability}
The Python programs that support the findings of this
study are openly available in the SPORT repository,
reference number \cite{FVides_SPORT}.

\section{Conclusion}
In conclusion, this study has demonstrated the effectiveness of equivariant autoregressive reservoir computers (EARCs) in identifying systems with inherent symmetries. Our comprehensive analysis revealed that EARCs can successfully capture the underlying dynamics of such systems, preserving their symmetrical properties. The use of sparse least-squares methods further enhanced the ability to discern approximate representations of the output coupling matrices, offering a novel approach to equivariant system identification. These findings not only deepen our understanding of equivariant systems but also open up new possibilities for the application of EARCs in various scientific and engineering domains. This work lays a foundational stone for future research, where the exploration of more complex symmetries and the integration of our methodologies into real-world scenarios could lead to interesting developments in the study and control of dynamical systems with symmetries.

\section*{Acknowledgment}
The computational simulations documented in this paper were performed with computational resources from the National Commission of Banks and Insurance Companies of Honduras (CNBS). The views expressed in the article do not necessarily represent the views of the National Commission of Banks and Insurance Companies of Honduras.

\end{document}